\documentstyle[psfig,aps,twocolumn]{revtex}

\newcommand{\rv}{{\bf r}}

\newcommand{\Bv}{{\bf B}}

\newcommand{\beq}{\begin{equation}}
\newcommand{\eeq}{\end{equation}}
\newcommand{\bea}{\begin{eqnarray}}
\newcommand{\eea}{\end{eqnarray}}

\newcommand{\<}{\langle}
\renewcommand{\>}{\rangle}

\begin{document}

\draft
\preprint{}
\title{Measurement of relative phase diffusion between two
Bose-Einstein condensates} 
\author{J. Ruostekoski and D. F. Walls} 
\address{ Department of Physics,
University of Auckland, Private Bag 92019, \\Auckland, New Zealand}
\date{\today}
\maketitle
\begin{abstract}
We propose a method of measuring diffusion of the relative phase between two
Bose-Einstein condensates occupying different nuclear or spin hyperfine states
coupled by a two-photon transition via an intermediate level. Due to the
macroscopic quantum coherence the condensates can be decoupled from the
electromagnetic fields. The rate of decoherence and the phase collapse may be
determined from the  occupation of the intermediate level or the absorption of
radiation.   
\end{abstract} \pacs{03.75.Fi,05.30.Jp,42.50.Dv}

In this paper we propose a method of measuring the diffusion of the relative 
phase between two Bose-Einstein condensates (BECs) formed in different
hyperfine states of the same atom. These two states are coupled by a
two-photon transition via an intermediate atomic level. Due to the macroscopic
quantum coherence of the BECs the two-photon transition between the two
hyperfine states is suppressed by quantum interference effects. This effect is
similar in origin to that occuring in electromagnetically induced transparency
(EIT) \cite{BOL91,MAR98} and lasing without inversion \cite{SCU91}. Atomic
interactions give rise to phase diffusion which destroy the coherence, so that
the two-photon transition is no longer completely suppressed. The rate of phase
diffusion may be determined by monitoring the population in the intermediate
level or the absorption rate.

Since the first realizations of dilute gas alkali BECs \cite{becexp} the
experiments have been broadened to include two and  multiple condensate
systems \cite{MYA97,STA98,MAT98,HAL98,HAL98b}. Myatt {\it et al.} \cite{MYA97}
produced two overlapping BECs of $|$F=1, m=-1$\rangle$ and $|$F=2, m=2$\rangle$
states of $^{87}$Rb using sympathetic cooling. The stability of this pair is due
to an unexpectedly small inelastic collision rate between these states
\cite{JUL97}. Recently, another BEC pair of $|$1, -1$\rangle$ and $|$2,
1$\rangle$ states of $^{87}$Rb has been realized at JILA
\cite{MAT98,HAL98,HAL98b}. These two states have essentially identical magnetic
moments and an adjustable spatial overlap allowing the creation of a fully
interpenetrating binary mixture. An especially interesting property is that
these two BECs can be coupled by a two-photon transition (one microwave and one
radiofrequency photon).

In general, inelastic collisions will limit the possibilities of magnetically
trapping BEC pairs. However, optical dipole traps, which use optical forces to
trap atoms, have a major advantage over magnetic traps since they can stably
trap atoms in arbitrary hyperfine states. An evaporatively cooled $^{23}$Na
gas has been succesfully confined in a dipole trap with a simultaneous 
observation of BECs in several different hyperfine states \cite{STA98}.

A fascinating property of BECs is that they exhibit a macroscopic quantum
coherence that is absent in thermal atomic ensembles \cite{coh}. Since one
needs a phase reference to observe a phase, binary mixtures of BECs are
especially useful in the studies of coherence properties. The atom-atom 
interactions in finite-sized BECs affect the matter wave coherence. The width
of the number distribution in the ground state has a dispersive effect on the
BEC self-interactions and the relative phase undergoes quantum collapses and
revivals \cite{WRI96,PAR98}. Additional sources of phase diffusion are spatial
mode fluctuations \cite{LEW96} and finite temperature decoherence due to the
interactions between condensate and noncondensate atoms \cite{JAK98,RUO98b}.

In this paper we consider a system closely related to the recent experiments 
of overlapping BECs in different hyperfine levels coupled by a two-photon
transition \cite{MAT98,HAL98,HAL98b}. Analogous BEC pairs could possibly be
produced also in dipole traps \cite{STA98}. As a consequence of the macroscopic
quantum coherence of BECs the two hyperfine levels coupled by a two-photon
transition exhibit two-photon coherence. By adjusting the initial conditions of
the BECs and the driving electromagnetic (EM) fields the atoms in the BECs can
be decoupled from the EM fields. However, as a consequence of the decoherence
and the collapse of the relative phase the two-photon coherence is reduced, and
the cw EM fields start inducing atomic transitions. This provides an excellent
scheme for measuring phase dynamics in the present experimental set-ups
\cite{MAT98,HAL98,HAL98b}. The two-photon coherence of two optically coupled
BECs has been previously predicted to result in various dramatic properties of
the scattered light \cite{JAV96b,SAV98}. Measurements of magnetic
coherence-related phenomena in atomic BECs have been previously addressed, e.g.,
in spin-polarized hydrogen \cite{SIG80} and in the electronic spin resonance of
alkali gases \cite{ZHA97}. 

We consider a three-level system with the energies of levels $|1\>$, $|2\>$, and
$|3\>$ denoted by $\omega_1$, $\omega_2$, and $\omega_3$. The microwave or rf cw
EM fields $\Bv_A$ and $\Bv_B$ induce magnetic dipole transitions between the
levels $1\leftrightarrow2$ and $2\leftrightarrow3$, respectively. For
simplicity, we  assume that the radiative lifetimes are long, so that the level
widths can be ignored. The EM interaction introduces the following terms into
the Hamiltonian density:  
\beq
{\cal H}_{\rm em}= - {\bbox \mu}_{12}\cdot\Bv_A \psi_1^\dagger\psi_2
- {\bbox \mu}_{23}\cdot\Bv_B \psi_2^\dagger\psi_3+{\rm H.c.} \,,
\eeq
where $\psi_i(\rv)$ is the field operator for hyperfine state $|i\>$ and ${\bbox
\mu}_{ij}$ denotes the magnetic moment for the transition $i\leftrightarrow j$.

We consider a situation where two macroscopic, perfectly overlapping BECs occupy
levels $|1\>$ and $|3\>$ and level $|2\>$ is initially empty. In
Refs. \cite{MAT98,HAL98,HAL98b}  the double BEC system in levels $|1\>$ and
$|3\>$ is prepared from the single BEC in state $|1\>$ by a two-photon
transition. If the occupation of level $|2\>$ is small, the collisional
interactions are mainly between atoms in levels $|1\>$ and $|3\>$. We write the
interaction Hamiltonian density in terms of the field operators $\psi_1(\rv)$
and $\psi_3(\rv)$ \beq
{\cal H}_{\rm int}=\sum_{i=1,3}{u_i\over2} \psi^\dagger_i 
\psi^\dagger_i \psi_i \psi_i + u_{13}\psi^\dagger_1
\psi^\dagger_3\psi_3 \psi_1 \,,
\label{eq:intham}
\eeq
where $u_i=4\pi\hbar^2 a_i/m$. We could directly obtain the equations of motion
for the condensate mean fields from Eq.~{(\ref{eq:intham})}, but the
interactions are significantly simplified in the case of approximately equal
scattering lenghts resulting in $u_1\simeq u_2\simeq u_{12}\equiv u$. In
$^{87}$Rb \cite{MAT98,HAL98}, the scattering lengths satisfy 
$a_1:a_{12}:a_2::1.03:1:0.97$, where $a_1$ ($a_2$) denotes the intraspecies
scattering length for state $|$1, -1$\rangle$ ($|$2, 1$\rangle$) and $a_{12}$ is
the interspecies scattering length. We approximate the field operators
$\psi_i(\rv) \simeq \phi_i(\rv)a_i$ for levels $|1\>$ and $|3\>$ in terms of the
BEC annihilation  operator $a_i$ and the corresponding spatial wave function
$\phi_i(\rv)$. It is also assumed that $\phi_1=\phi_3\equiv\phi$. Our aim is to
write the equations of motion for the expectation values $\sigma_{ij}\equiv 
\<a_i^\dagger a_j\> /N $, where the total initial number of BEC atoms is denoted
by $N$. The effect of noncondensate atoms  is treated by a phenomenological
damping in the equation of motion for the coherence between the two BECs. By
approximating the scattering lengths to be equal and by assuming $\<a_2^\dagger
a_2\> \ll N$ we may then approximate the interaction Hamiltonian
[Eq.~{(\ref{eq:intham})}] by     
\beq 
H_{\rm int}\simeq
{\hbar\kappa\over2}(\hat{N}^2-\hat{N}-2\hat{N}a_2^\dagger a_2)  \,,
\label{eq:intham2}
\eeq
where $\hat{N}=\sum_i a_i^\dagger a_i$ is assumed to be a constant of the motion
and $\hbar\kappa=u\int d^3r|\phi|^4$. We see that for perfectly overlapping BECs 
with approximately equal scattering lengths the effect of BEC
self-interactions is strongly suppressed. Finally, the detunings are defined by
$\delta_{21}\equiv\omega_A-\omega_{21}$ and
$\delta_{32}\equiv\omega_B-\omega_{32}$, where $\omega_A$ ($\omega_B$) is the
frequency of the field $\Bv_A$ ($\Bv_B$) and the transition frequency between
the hyperfine levels $1\leftrightarrow2$ ($2\leftrightarrow3$) is $\omega_{21}$
($\omega_{32}$). In the rotating-wave approximation we then obtain the following
equations of motion for the expectation values $\sigma_{ij}$:             
\begin{mathletters}   
\bea 
\dot{\sigma}_{11} &=& \Omega_A {\rm
Im}(\sigma_{21}),\\ 
\dot{\sigma}_{22} &=& -\Omega_A {\rm Im}(\sigma_{21})+\Omega_B  {\rm
Im}(\sigma_{32}),\label{d22}\\  
\dot{\sigma}_{33} &=& -\Omega_B {\rm Im}(\sigma_{32}),\\
\dot{\sigma}_{21} &=& -i(\delta_{21}+N\kappa) \sigma_{21} +{i\Omega_A\over2}
(\sigma_{22}-\sigma_{11})-{i\Omega_B\over2} \sigma_{31}, \label{d21}\\
\dot{\sigma}_{32} &=& -i(\delta_{32}-N\kappa) \sigma_{32} +{i\Omega_B\over2}
(\sigma_{33}-\sigma_{22})+{i\Omega_A\over2} \sigma_{31}, \label{d32}\\
\dot{\sigma}_{31} &=& -i(\delta_{32}+\delta_{21}-i\gamma) \sigma_{31}
+{i\Omega_A\over2} \sigma_{32}-{i\Omega_B\over2}  \sigma_{21} ,
\eea
\label{eq:sys}
\end{mathletters}
where the Rabi frequencies are given by $\Omega_A\equiv 2\int d^3r \phi^*_2 
\phi_1 {\bbox \mu}_{21}\cdot\Bv_A/\hbar $, $\Omega_B\equiv 2\int d^3r
\phi^*_2  \phi_3 {\bbox \mu}_{23}\cdot\Bv_B/\hbar $, and Im denotes  the
imaginary part. For simplicity, in Eq.~{(\ref{eq:sys})} we have set $\Omega_A$
and $\Omega_B$ to be real. Here the equations correspond to the cascade or
ladder three-level system with $\omega_1<\omega_2<\omega_3$. In the case of
$\Lambda$ (V) three-level scheme the sign of $\delta_{32}$ ($\delta_{21}$)
should be changed. The dominant mean-field contribution of the BEC
self-interactions is to shift the resonance conditions in Eqs.~{(\ref{d21})} and
{(\ref{d32})}.

Level $|2\>$ is assumed to be initially empty and two BECs occupy levels $|1\>$
and $|3\>$. The off-diagonal element $\sigma_{31}$ describes the macroscopic
coherence between the two BECs. As explained earlier this collapses and
decoheres due to the atom-atom interactions. We have included the effect of the
decay of the matter wave coherence in Eq.~{(\ref{eq:sys})} in terms of a
phenomenological damping parameter $\gamma$ in the equation of motion for
$\sigma_{31}$. This damping parameter includes contributions from both the
quantum effects of BEC self-interactions and collisions between condensate and
noncondensate atoms. We assume that these dominate over other damping mechanisms
as long as the population in level $|2\>$ remains small.

For an initial condition for Eq.~{(\ref{eq:sys})} we set
$\sigma_{11}=\sigma_{33}=\sigma_{31}=1/2$ indicating a well-established 
coherence between the two BECs with a vanishing relative phase. This
corresponds, e.g., to a situation where a BEC is first prepared in level $|1\>$
and half of the BEC atoms are then coherently transferred to level $|3\>$, so
that the atoms remain entangled. Initially there are no atoms in the 
intermediate level $\sigma_{22}=0$. The EM fields are two-photon resonant
($\delta_{21}=-\delta_{32}$) and $\Omega_A=-\Omega_B\equiv\Omega$. It is easy to
see from Eq.~{(\ref{eq:sys})} that for $\gamma=0$ this corresponds to a
steady-state situation. Both BECs are decoupled from the EM fields and the
absorption described by Im$(\sigma_{21})$ and Im$(\sigma_{32})$ vanishes
\cite{com}. However, for non-zero $\gamma$, Im$(\sigma_{21})$ and
Im$(\sigma_{32})$ also become non-zero, and atoms start accumulating in level
$|2\>$ due to the absorpion of EM radiation. The phase damping parameter
$\gamma$ may be determined by measuring the oscillations of the EM fields or the
population in level $|2\>$.

Equations (\ref{eq:sys}) may be integrated numerically, but it is also
illuminating to look at analytic estimates. We consider a situation where the
EM fields are resonant, $\delta_{21}+N\kappa=\delta_{32}-N\kappa=0$, and the
damping $\gamma$ is much smaller than the Rabi frequency $\Omega$. We look for 
an exponential solution to Eq.~{(\ref{eq:sys})} to leading order in the small
parameter $\gamma/\Omega$. By also determining the coefficients to first order 
in $\gamma/\Omega$ we obtain $\sigma_{31}(t)$, describing the matter wave
coherence between the BECs, as 
\beq
\sigma_{31}(t)\simeq {1\over2}\,e^{-3\gamma t/4}-{\gamma\over 8\sqrt{2}\Omega}\,
e^{-\gamma t/8} \sin{(\sqrt{2}\Omega t)}\,.
\eeq
The absorption of the field $\Bv_A$ is proportional to
\beq
{\rm Im}[\sigma_{21}(t)] \simeq -{\gamma\over8\Omega}\, e^{-3\gamma
t/4}+{\gamma\over 8\Omega}\, e^{-\gamma t/8} \cos{(\sqrt{2}\Omega t)}\,.
\label{eq:21}
\eeq
The absorption results in oscillating EM fields with the amplitude of the
oscillations given by $\gamma/(8\Omega)e^{-\gamma t/8} $. The real parts satisfy
Re$[\sigma_{21}(t)]=$Re$[ \sigma_{32}(t)] =0$. Finally, the occupation in
level $|2\>$ is 
\beq
\sigma_{22}(t)\simeq {1\over3} (1- e^{-3\gamma t/4})-{\gamma\over
4\sqrt{2}\Omega} \, e^{-\gamma t/8} \sin{(\sqrt{2}\Omega t)}\,.
\label{eq:22}
\eeq 
Due to the decoherence of the BECs the EM fields absorb radiation and atoms
start occupying level $|2\>$. By measuring the number of atoms in state
$|2\>$ at time $t$ after switching on the driving EM fields, one could
determine the damping rate of the matter wave coherence $\gamma$.
Alternatively, the damping rate could be observed from the amplitude of the
oscillating EM signal.

\begin{figure}
\begin{center}
\leavevmode
\psfig{bbllx=1cm,bblly=7cm,bburx=19cm,bbury=15cm,width=7cm,file=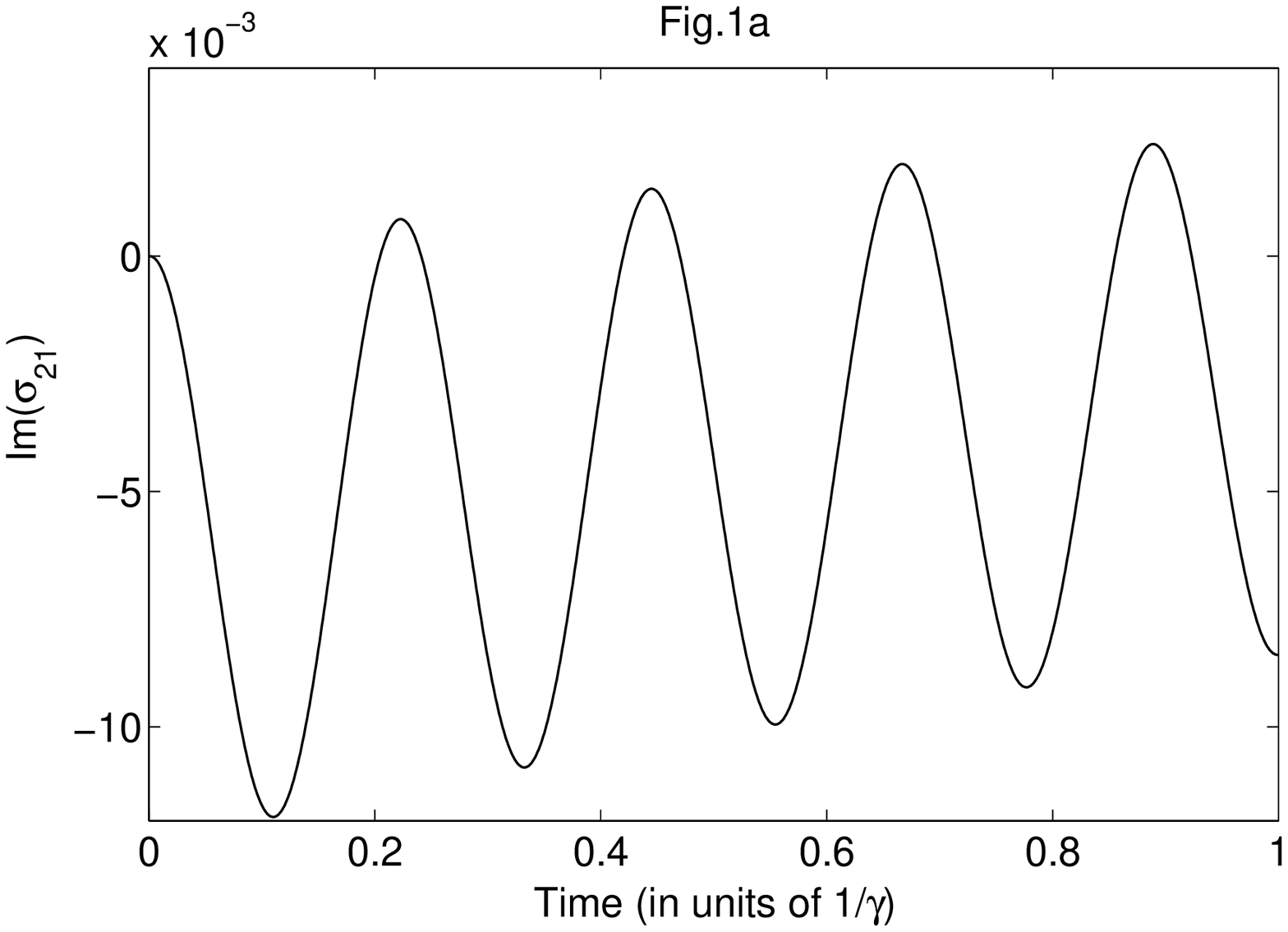}\vspace{2cm}
\psfig{bbllx=1cm,bblly=7cm,bburx=19cm,bbury=15cm,width=7cm,file=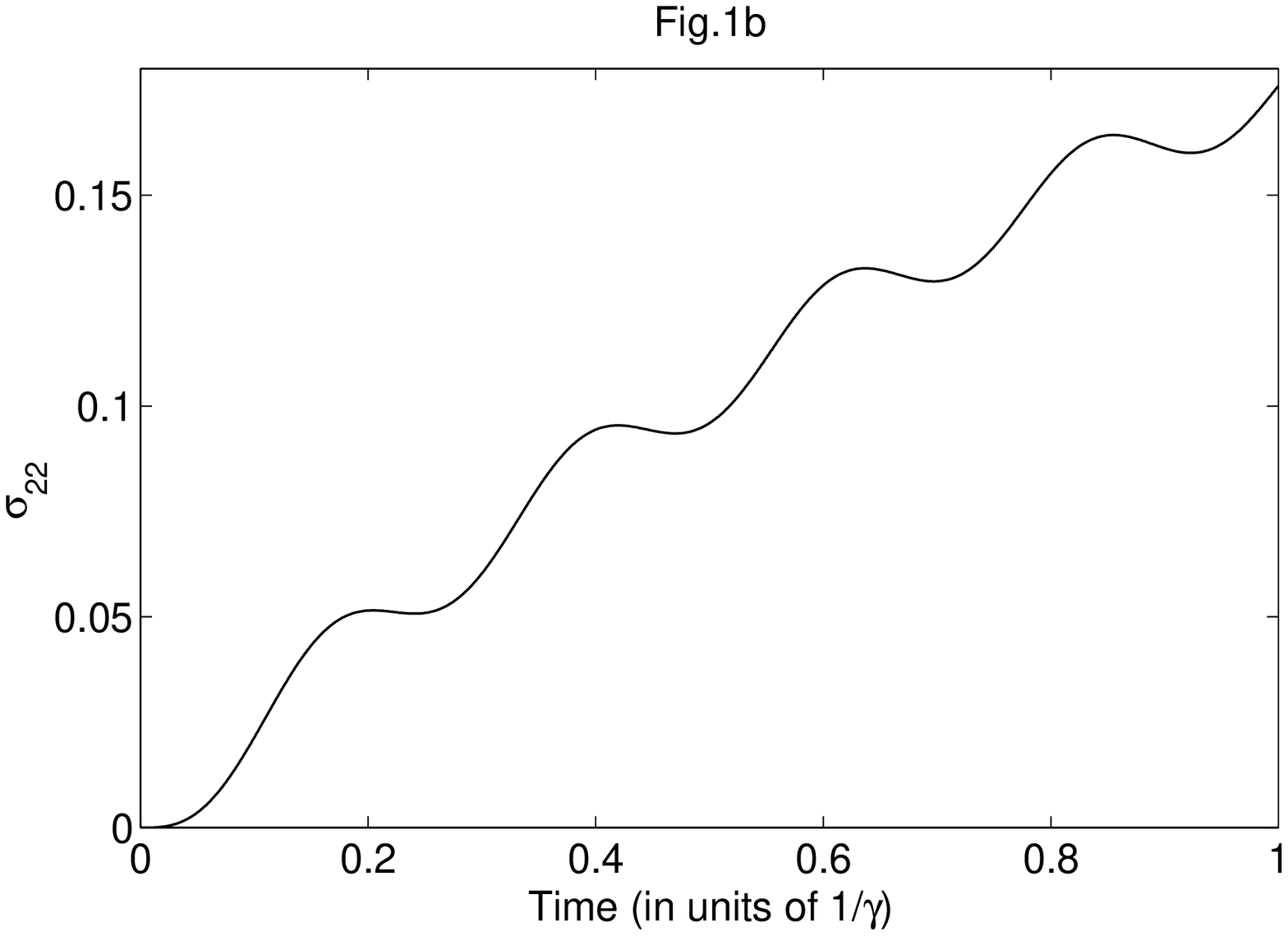}
\end{center}
\caption{A particular example response of the medium in the case of
$\Omega=20\gamma$. (a) ${\rm Im}[\sigma_{21}(t)]$ proportional to the absorption
of the EM radiation from the field driving the transition between levels $|1\>$
and $|2\>$. (b) $\sigma_{22}(t)$ proportional to the population of the
intermediate level. The EM fields are assumed to be resonant.
}   
\label{fig:1}   
\end{figure}

In Fig.~{\ref{fig:1}} we have plotted one example of the signal corresponding
to a particular value $\Omega=20\gamma$. We plot ${\rm Im}[\sigma_{21}(t)]$ and
$\sigma_{22}(t)$ obtained by numerically integrating Eq.~{(\ref{eq:sys})}
for  $\delta_{21}+N\kappa=\delta_{32}-N\kappa=0$, and for the initial condition
$\sigma_{11}(0)=\sigma_{33}(0)=\sigma_{31}(0)=1/2$ and
$\sigma_{21}(0)=\sigma_{32}(0)=0$. The oscillating signal (a) and the
accumulating population in the intermediate level (b) are clearly observed. The
graphs are also well represented by the approximate analytic results,
Eqs.~{(\ref{eq:21})} and {(\ref{eq:22})}.

\vspace{2cm}

If the fields are off-resonant from intermediate level $|2\>$ the absorption of
the radiation and the occupation of state $|2\>$ are reduced, but ${\rm
Re}(\sigma_{21})$ and ${\rm Re}(\sigma_{32})$ are in this case non-zero.

In EIT the interference between the off-diagonal density matrix elements leads
to an initially opaque medium being rendered almost transparent
\cite{BOL91,MAR98}. The transition amplitudes driven by two oppositely phased EM
fields destructively interfere. The effect of the medium on the EM fields is
canceled. The crucial quantities for the interference are the coherences
$\sigma_{21}$ and $\sigma_{32}$ describing the transition dipole matrix elements
excited by the driving EM fields. In EIT the coherence between levels $|1\>$ and
$|3\>$ is present only as a consequence of the EM coupling via level $|2\>$. For
BECs in hyperfine states $|1\>$ and $|3\>$ the coherence $\sigma_{31}$ is 
present from the start without being created by the driving fields due to the
macroscopic matter wave coherence. As a result, the EM fields couple
$\sigma_{31}$ to the oscillating dipoles. The presence of the macroscopic
quantum coherence of BECs can then completely inhibit the EM fields from
establishing any coherence between levels $|1\>$ and $|2\>$ or $|2\>$ and 
$|3\>$. On the other hand, the rate of which coherences $\sigma_{21}$ and
$\sigma_{32}$ are induced describes the decoherence rate of BECs.

In recent experiments Hall {\it et al.} \cite{HAL98b} studied phase diffusion
in a binary BEC mixture of $^{87}$Rb. A BEC was first prepared in level $|$1,
-1$\rangle$ and a part of the condensate was then transferred to level $|$2,
1$\rangle$. The relative phase between the two separated halves was determined
by interfering the atoms at a later time. The phase diffusion rate was estimated
by varying the evolution time of the two BECs before the interference
measurement. In every interference measurement the BECs were destructively
imaged and the repetitions of independent runs produced information about the
uncertainty of the phase. Only weak phase diffusion was observed. 

In addition to the enviromentally-induced decoherence due to the interactions
between condensate and thermal noncondensate atoms \cite{JAK98,RUO98b} the
quantum collapse due to the BEC self-interactions has an important effect on 
the phase diffusion \cite{WRI96,PAR98}. This rate dramatically depends on the
relative strength of the three scattering lengths $a_1$, $a_3$, and $a_{13}$ in
Eq.~{(\ref{eq:intham})}. Under conditions where the scattering lengths are
equal, the two BECs are perfectly overlapping, and only levels $|1\>$ and
$|3\>$ are occupied, the interaction Hamiltonian in Eq.~{(\ref{eq:intham2})}
depends only on the constant total atom number. This suppresses the phase
collapse.

Recent experiments have realized overlapping BECs in different hyperfine states
\cite{MYA97,STA98,MAT98,HAL98,HAL98b}. A two-photon transition between a BEC pair
of $|$1, -1$\rangle$ and $|$2, 1$\rangle$ states of $^{87}$Rb via level $|$2,
0$\rangle$ has been implemented \cite{MAT98,HAL98,HAL98b}. The intermediate
state $|$2, 0$\rangle$ of $^{87}$Rb for the two-photon coupling is untrapable.
The atoms in $|$2, 0$\rangle$ can escape the trap. This would correspond in our
scheme to an additional damping $\Gamma$ for level $|2\>$. If the population in
level $|2\>$ is small, this damping would be approximately independent of the
number of atoms in $|2\>$. In that case we could add the following additional
term to Eq.~{(\ref{d22})}:  $\dot{\sigma}_{22}=\ldots -\Gamma\sigma_{22}$. Then
the phase diffusion could possibly be measured by monitoring the number of atoms
escaped through level $|$2, 0$\rangle$ or by counting the atoms remained in
levels $|$1, -1$\rangle$ and $|$2, 1$\rangle$.  Optical dipole traps
\cite{STA98} can stably trap atoms in arbitrary hyperfine levels. Suitable BEC
pairs, without losses of atoms, could possibly be produced in dipole traps
to implement the proposed scheme for the measurement of the phase diffusion.

In conclusion, we have proposed a method of measuring the ``phase memory" of a
BEC pair. This method relies on the quantum interference of transition
amplitudes and is similar in origin to that occuring in EIT and lasing without
inversion. Unlike the previous measurements of phase diffusion \cite{HAL98b},
our model allows continuous and nondestructive monitoring of the phase dynamics.

We would like to thank M.\ J.\ Collett for helpful comments.
This work was supported by the Marsden Fund of the Royal Society of New Zealand
and The University of Auckland Research Fund.


\begin{references}
\bibitem{com} To decouple the BECs from the EM fields in the case of a
nonvanishing relative phase between the BECs, $\Delta\varphi \equiv {\rm arg}(\<
a^\dagger_3 a_1\> )$, we need to choose 
$\Delta\phi=\pi+\Delta\varphi$, where $\Delta\phi\equiv {\rm arg}(\Omega_B^*
\Omega_A )$ is the phase difference between the Rabi frequencies.
\end{references}
\end{document}